\documentclass[grl]{agujournal2019}
\usepackage{url} 
\usepackage{lineno}
\usepackage[inline]{trackchanges} 
\usepackage{soul}
\usepackage{amstext}
\usepackage{amssymb}
\usepackage{ragged2e} 

\draftfalse
\journalname{Geophysical Research Letters}

\begin{document}
\justifying

\title{Rare event algorithm study of extreme warm summers and heatwaves over Europe}

\authors{F. Ragone\affil{1,2,3} and F. Bouchet\affil{1}}
\affiliation{1}{Univ Lyon, Ens de Lyon, Univ Claude Bernard, CNRS, Laboratoire de Physique, Lyon, France}
\affiliation{2}{Earth and Life Institute, Université catholique de Louvain, Louvain-la-Neuve, Belgium}
\affiliation{3}{Royal Meteorological Institute of Belgium, Brussels, Belgium}

\correspondingauthor{Francesco Ragone}{francesco.ragone@uclouvain.be}


\begin{keypoints}
\item The rare event algorithm increases by several orders of magnitude the number of warm summers and heatwaves sampled by the model.
\item Warm summers over either France or Scandinavia are linked to wavenumber 3 hemispheric teleconnection patterns.
\item Warm summers in Scandinavia show bimodality due to different distribution of subsequent subseasonal heatwaves.
\end{keypoints}

\begin{abstract}
The analysis of extremes in climate models is hindered by the lack of statistics due to the computational costs required to run simulations long enough to sample rare events. We demonstrate how rare event algorithms can improve the statistics of extreme events in state-of-the-art climate models. We study extreme warm summers and heatwaves over France and Scandinavia with CESM1.2.2 in present-day climate. The algorithm concentrates the simulations on events of importance, and shifts the probability distributions of regional temperatures such that warm summers become common. We estimate return times of extremes orders of magnitude larger than what feasible with direct sampling, and we compute statistically significant composite maps of dynamical quantities conditional on the occurence of the extremes. We show that extreme warm summers are associated to wavenumber 3 hemispheric teleconnection patterns, and that the most extreme summers are related to the succession of rare subseasonal heatwaves.
\end{abstract}

\section*{Plain Language Summary}
The impact of extreme climatic events is often dominated by the rarest events. These events have return times (a measure of how often they occur on average) of hundreds of years or more, but they could, and do, happen anytime. These events are poorly understood because of lack of statistics. Climate models are computationally expensive, and can not be run for long enough to study events with return times longer than a few decades. We use a new computational technique that allows simulations to focus only on trajectories leading to extreme heatwaves over a target region, optimizing the use of computational resources. We thus gather robust statistics for seasonal heatwaves with return times of hundreds or thousands of years, and observe even rarer events impossible to observe otherwise. We find that extreme warm summer over France or Scandinavia are synchronised with extreme warm summers in specific regions of Asia and North-America by a teleconnection pattern extending over the entire Northern hemisphere. We also find suggestions that extreme warm summers over Scandinavia may occur in two distinct ways. This new method can be used to better study the impact of global warming on the risk of catastrophic events, and to improve their predictability.

\section{Introduction}\label{introduction}

Recent decades have seen a number of exceptionally warm summers and record breaking heatwaves at the Northern hemisphere midlatitudes \cite{Luterbacher2004,garcia-herrera_review_2010,Barriopedro_2011,Otto_2012,AghaKouchak2012,IPCC_2012,Russo_2015,coumou_decade_2012}. The Intergovernmental Panel on Climate Change (IPCC) has concluded that hot days and heavy precipitation events have become more frequent since 1950 \cite{IPCC_2013}. The precise extent to which heat extremes will become more common in the future and under the target scenarios of +1.5 $^o$C and +2 $^o$C global surface temperature with respect to preindustrial levels is an active field of research \cite{Mueller2016,Dosio_et_al_2018,Suarez_Gutierrez_2018}. 

A phenomenological theory exists on the causes of heatwaves \cite{schubert_northern_2014,Perkins2015,Horton2016}, including large scale atmospheric circulation patterns, \cite{della-marta_summer_2007,Cassou2005,Jezequel_et_al_2018}, lack of precipitation and soil moisture \cite{Vautard&al2007,Zampieri&al2014,DAndrea&al2016}, and sea surface temperature anomalies \cite{della-marta_summer_2007,Cassou2005}, with different levels of importance in different regions \cite{Stefanon_2012}. Long lasting heatwaves and warm summers are among the most relevant events in terms of impacts \cite{Camargo2016,Trenberth2012}. Although they are known to be related to persistent weather regimes \cite{LauKim_2012,Branstator_2013,Hoskins2015,Petoukhov_2016,Boers2019,Kornhuber_2019}, their dynamics is not  fully understood. Some studies show stronger increase with global warming of the frequency of persistent temperature extremes \cite{Pfleiderer&al2019}. However, the response of the global circulation to climate change is extremely complex and not well understood \cite{Hoskins2015}, in particular for the type of rare dynamics that lead to extreme events. 

The impact of climate extremes is often dominated by the rarest events. For instance, the death toll of the Western European heatwave in 2003, about 70.000 casualties, exceeded the sum of the fatalities due to all other heatwaves during the last three decades \cite{Barriopedro_2011}. Moreover, extreme events with very long return times (also called return periods) always occur. For instance, an extreme event with a return time of 1000 years, has a chance 1/1000 to occur next year. Among the huge number of possible extreme events with a return time of 1000 years, a few of them will certainly occur somewhere sometimes next year. It is thus crucial to study the most extreme and rare events However, the study of such events faces strong scientific limitations: we can not rely on historical data for events with return times of 100 years or more, because often no similar events have ever been observed, and it is extremely difficult to study these events using climate models because of the required computational costs.

In this paper we address the problem of computational cost limitations. In order to face this problem, we use a rare event algorithm that concentrates ensemble simulations on trajectories of importance for the extreme events, optimizing the use of the computational resources. Rare event algorithms have recently been applied to turbulence problems
\cite{Grafke,Laurie,Grauer,Bouchet_Rolland_Simonnet_2019:C,lestang2020numerical} and climate applications \cite{Ragone&al2018,webber_practical_2019,Ragone_Bouchet2020,plotkin2019maximizing}. 
Based on the phenomenology of the dynamics for each family of extreme events, an appropriated type of rare event algorithm should be carefully chosen. Here we use a genealogical algorithm \cite{Ragone&al2018,Ragone_Bouchet2020} adapted from \cite{del2005genealogical,giardina_simulating_2011}, that is efficient to study long lasting events. This rare event algorithm has been used to study heatwaves with an intermediate complexity model \cite{Ragone&al2018} in perpetual summer conditions. Here we use it to sample rare heatwaves in CESM 1.2.2 \cite{CESM}, a fully realistic climate model, in presence of daily and seasonal cycles. 

The goal of this paper is to demonstrate the applicability of rare event algorithms to state-of-the-art climate models, and to highlight properties of extreme long lasting heatwaves that can not be assessed with traditional sampling strategies. We study extreme warm summers and heatwaves over France and Scandinavia in CESM1.2.2, in present-day climate. We compute return times orders of magnitude larger than what feasible with direct sampling, and statistically significant composite maps of dynamical quantities. Our results show that extreme warm summers are associated to wavenumber 3 teleconnection patterns in the Northern hemisphere, and  suggest that the most extreme summers are related to the succession of multiple rare subseasonal heatwaves.

\section{Data and methods}\label{sec:data_methods}


{\bf Model.} All simulations are performed with the Community Earth System Model (CESM) version 1.2.2 \cite{CESM}. We use an atmosphere and land only setup, whose active components are the Community Atmospheric Model version 4 (CAM4) and the Community Land Model version 2 (CLM2). The model is run at statistically stationary state, with sea surface temperature (SST), sea ice cover, and the concentration of atmospheric CO2 and other greenhouse gases fixed at values representative of present day climate (year 2000). Contrary to \cite{Ragone&al2018}, the model features daily and seasonal cycle, and reproduces a fully realistic climate. See the SI for more details. We study the statistics of a $1000$ years long control run, and the statistics of several sets of simulations using the rare event algorithm. 

{\bf Heatwaves and physical observable definition.} We consider the surface temperature $T_{s}(\vec{r},t)$, where $\vec{r}$ is the space variable and $t$ is time, and define the mean surface temperature $\mathbb{E}\left(T_{s}\right)(\vec{r},t)$, which varies in time (because of the seasonal cycle) and in space. In practice, $\mathbb{E}\left(T_{s}\right)$ will be approximated by the climatological average of $T_{s}$ computed from a $1000$ years long control run. We study the statistics of the spatially and temporally averaged surface temperature anomaly defined by  
\begin{equation}
\label{definition_A}
a(t_I)=\frac{1}{T}\int_{t_I}^{t_I+T} A(t) \mathrm{d}t \ \ \ {\rm where } \ \ \ A(t) = \frac{1}{\mathcal{\left|D\right|}}\int_{\mathcal{D}}\left(T_{s}-\mathbb{E}\left(T_{s}\right)\right)(\vec{r},t)\,\mathrm{d}\vec{r},
\end{equation}
where $A(t)$ is the instantaneous spatial average, $\mathcal{D}$ is a spatial domain area, $t_I$ is the starting date of the time average, $T$ is the averaging time, for instance several days up to one season. Heatwaves will be defined as extreme values of the observable $a(t_I)$, as heatwave indices developed for dynamical studies often use anomalies rather than absolute values \cite{Perkins2015}. We are specifically interested in long events, for instance warm summers with $T$=3 months. For summer anomalies, we consider $a_{JJA}$ defined by  (\ref{definition_A}) with $t_I$ being June 1st and $t_I+T$ August 29. In the following, we consider ${\mathcal D}$ as the area over either France or Scandinavia (see the SI for the definition of the areas).


{\bf Rare event algorithm experiments and importance sampling.} Rare event algorithms allows a numerical model to produce very efficiently rare trajectories of importance for the study of extremes. We simulate an ensemble of $N$ model trajectories, and at constant intervals of a resampling time $\tau$ we perform trajectory selection, pruning and cloning trajectories in order to favour those leading to the extreme events. See the SI for a more detailed description and \cite{Ragone&al2018}. The algorithm also allows to compute probabilities for the obtained trajectories. Let $\vec{X(t)}$ be the vector of values of all the model variables at time $t$ (which includes but is not limited to the temperature $T_{s}$ defined above). We consider the trajectory $\left\{\vec{X}(t)\right\}$, with all values of $X$ along all the interval $t_a\leq t \leq t_a+T_a$, where $t_a$ is the starting date and $T_a$ the trajectory duration in the algorithm. We denote $\mathbb{P}_k\left(\left\{\vec{X}(t)\right\}\right)$ the probability distribution function of the trajectories obtained in the algorithm. Then the rare event algorithm produce trajectories distributed according to
\begin{equation}
\label{tilting}
\mathbb{P}_k\left(\left\{\vec{X}(t)\right\}\right)=\frac{e^{k\int_{t_{a}}^{t_{a}+T_a} A(u)\,\mathrm{d}u}}{Z}\mathbb{P}_0\left(\left\{\vec{X}(t)\right\}\right),
\end{equation}
where $\mathbb{P}_0\left(\left\{\vec{X}(t)\right\}\right)$ is the probability distribution function of the trajectories in the model climate, $Z$ is a normalization term computed by the algorithm, $A$ is the selection function (with the dependance on the trajectory implicit for simplicity, $A(t)=A\left(\left\{\vec{X}(t)\right\}\right)$), and $k$ controls the strength of the algorithm selection. Equation \ref{tilting} is called an importance ratio formula. Using equation \ref{tilting} we can compute the actual probabilities  of the rare trajectories obtained in the simulations with the algorithm. See the SI and \cite{Ragone&al2018} for more details.

We use as selection function $A$ the same function used to define a heatwave in formula (\ref{definition_A}). We see from equation (\ref{tilting}) that if the parameter $k$ is positive, then trajectories with large values of the time average of the surface temperature anomaly will be much more probable in simulations obtained with the algorithm rather than in simple simulations with the model. The larger $k$, the stronger the selection, and the more probable trajectories with extreme values of the time averaged surface temperature. We analyse extremely warm summers over France and Scandinavia. For each case, we perform $K=10$ ensemble simulations with the algorithm, each with $N$=100 trajectories, biasing parameter $k$=30, with $T_a$=90 days from June 1st to August 29th, and resampling time $\tau$=5 days (see the SI). A 1000 years long control run is used to provide initial conditions for the experiments with the algorithm, and as a benchmark for the statistics. The computational cost of the experiments with the algorithm is equivalent to simulating 1000 summers in the control run, but they allow to gather a much richer statistics for the extreme events of interest.

\section{Results}\label{results}
\subsection{Importance sampling of extreme warm summers}\label{results_algorithm_sampling}

The main goal of the algorithm is to perform importance sampling for the distribution of the summer temperature anomalies over the target region. Figures 1a and 1b show the probability distribution functions of the summer temperature anomalies $a_{\mbox{\tiny JJA}}$ for the control run and the rare event algorithm. The control distribution shows a similar variance of about 1 $^o$K for both France and Scandinavia. The algorithm is very effective in performing importance sampling and populating the upper tails of the distributions. The typical value of the seasonal anomaly $a_{JJA}$ in the rare event algorithm experiments is around 4 $^o$K for both cases, which are values never observed in the control run. Thanks to the algorithm most of the computational power is indeed used to simulate extremely warm summers, rather than trajectories belonging to the bulk of the distribution. In the case of Scandinavia the algorithm statistics in figure 1b shows a bimodality that we will discuss in section \ref{results_subseasonal}.

In figures 1c,d we compare return times of $a_{JJA}$ in the control run and in the experiments with the rare event algorithm. A description of the computation of return times both for direct sampling and importance sampling using formula (\ref{tilting}) is presented in \cite{Lestang_2018}. The black curves are obtained from the control run using 1000 years of data. In order to estimate uncertainty ranges, we compute also an estimate dividing the control run in $K$=10 samples of 100 years each, computing 10 estimates of the return times curve. We then take their average (blue curves), and compute the empirical standard deviation (blue shaded areas). The red line and shaded areas are obtained in the same way, but using the estimates from $K=10$ independent experiments with the algorithm. 

With the rare event algorithm we reach return times up to 10$^5$-10$^6$ years with uncertainty ranges comparable with the ones with the control run for return times of order 10$^2$ years. Given the large value of $k$ we chose, we have a small range of overlap for estimates of the return times from the control run and the rare event algorithm. When they do overlap, the values are consistent with each other within the uncertainty ranges.  After about 10$^6$-10$^7$ years the return time curves reach a plateau. Such plateaux are due to undersampling, as discussed in the SI and in \cite{Lestang_2018}. A full demonstration of the reliability of the results obtained with the rare event algorithm in similar simulations can be found in \cite{Ragone&al2018,Ragone_Bouchet2020}. 

\subsection{Teleconnection patterns for extreme warm summers}\label{results_algorithm_teleconnections}

We study the dynamical properties of extremely warm summers with return time larger than 100 years (called 100-year warm summers or seasonal heatwaves from now on). We compute composite maps of anomalies of the local JJA surface temperature and 500 hPa geopotential height conditional on the occurrence of 100-year warm summers, for the control run and the rare event algorithm experiments. One of the key advantage of the rare event algorithm is that it gives much better results than the control run for composite statistics for large return times. Moreover the rare event algorithm gives access to composite statistics for return times larger than $1000$ years, which is impossible with the control run.

Figure 2a shows composite statistics of surface temperature and 500 hPa geopotential height of 100-year warm summers over France in the control run. The map hints at the presence of a wavenumber 3 teleconnection pattern for both observables. However, how much of these patterns, which were obtained by averaging over only 10 warm summers, is statistically significant? Figure 2b shows the $t$-value map for the geopotential height (see the SI for details). Only the geopotential height anomalies over Europe pass a statistical significance test with $|t|>2$, which means then we can not really assess the reality of the teleconnection pattern using a 1000 years control run: we just do not have enough data. Figures 2c,d show the same composite maps and statistical significance analysis for 100-year warm summers over France, but computed using the rare event algorithm. The algorithm results are globally significant, except over the northern part of the Pacific area. Transition areas between cyclonic and anticyclonic anomalies have a $|t|$ value smaller than $2$, because they have a low value of the conditional average. They are however rather narrow, so that their location can be still considered rather precise. The rare event algorithm thus allows to properly assess the existence of teleconnections of warm summers in Europe, North America and Central Asia.

The algorithm also gives a much better estimate of the amplitude of the anomalies, since it gives access to large amplitude events unavailable in the control run, because they are too rare. In the same way, the rare event algorithm gives also access to composite statistics that  totally unavailable with the control run. For instance figure 3a shows composite statistics of surface temperature and 500 hPa geopotential height for 1000-year warm summers over Scandinavia, which also show a teleconnection pattern with a different spatial pattern but similar broad stroke features.

Based on the rare event algorithm results, we obtain a better overview of the characteristics of the synoptic dynamics occurring during extremely warm summers in the considered regions. A warming pattern centered over the target region is present, encompassing a larger area on a spatial scale of a few thousands km, which coincide with central-Western Europe for France and Northern-Eastern Europe for Scandinavia. Persistent anticyclonic synoptic scale structures centered over the target area are associated to this warming pattern. Locally these structures are consistent with the observed  synoptic conditions for the occurrence of European heatwaves. In particular, for the North Atlantic and European area, the local patterns are very close to the observed patterns for the Western-European and the Scandinavian heatwave clusters obtained from reanalysis data in \cite{Stefanon_2012}. 

The application of the rare event algorithm allows to assess in a statistically robust way that these local dynamics are part of hemispheric structures of approximately wavenumber 3, which induce a  temperature teleconnection pattern, with persistent regional temperature anomalies of alternating signs around the hemisphere. The patterns for the two regions are in broad strokes similar, although they differ in the exact location of positive and negative anomalies. Warm summers over France occur systematically with warm summers over Siberia and North-East America. The corresponding tripolar structure of anticyclonic anomalies is accompanied by a localized low over Central-Asia and a general lower pressure over the Arctic, with minimum over Greenland. The structure related to Scandinavian warm summers is similar on the North-Atlantic sector, but over Asia it is quite different, with a negative temperature anomaly (and related low pressure) extending from Southern Europe to the whole central Asia, with the positive anomaly constrained over the far East. The circulation over the Arctic is also different, with a strong low over the Pole. 

\subsection{Bimodality of warm Scandinavian summers and subseasonal heatwaves}\label{results_subseasonal}

Figure 1b a bimodality of the Scandinavia summer temperatures in the rare event algorithm data. This suggests that two types of distinct dynamical events might lead to Scandinavian extreme summers. We note that, because of the nonlinear relation (\ref{tilting}), a bimodality in the algorithm distribution does not necessarily imply a bimodality in the tail of the model distribution. Still it is interesting to test the hypothesis of two types of dynamics. 

Figure 3a shows composite statistics of surface temperature and 500 hPa geopotential height for 1000-year warm summers over Scandinavia. Figures 3b and 3c show the composite maps from the rare event algorithm computed separately for $a_{\mbox{\tiny JJA}}<4.2 ^oK$ and $a_{\mbox{\tiny JJA}}>4.2 ^oK$, where 4.2 $^o$K corresponds to the local minimum between the two peaks of the distribution in red in figure 1b. While the overall structure is the same as the total composite map in figure 3a, the map of the first range (figure 3b) shows a weaker teleconnection pattern, and an anticyclonic anomaly over the North-Atlantic that is not present in the map of the second range (figure 3c). The Scandinavian warm summer dynamics is compatible with a northward shift of the jet stream along the entire hemisphere. In the case of the first range (figure 3b), this shift is less clear over the North-Atlantic, where a different dynamics seems to be in place. It is however likely that, if two distinct types of dynamics are occurring, a selection based only on the range of seasonal regional temperature fluctuations is not enough to differentiate between different dynamics, and therefore simple composite maps show from that point of view mixed information. 

The study of relation between subseasonal fluctuations of surface temperature and extreme warm summers can give an explanation of this bimodality. To visualize this relation, we look at the genealogical structure of the trajectories in an experiment with the rare event algorithm for Scandinavia in figure 4a. The black lines correspond to trajectories belonging to the first range ($a_{\mbox{\tiny \rm JJA}}<4.2^o$K), while the red lines to trajectories belonging to the second range ($a_{\mbox{\tiny \rm JJA}}>4.2^o$K). In the experiment represented in figure 4a, trajectories belonging to each of the two ranges of the bimodal distribution coexist. In figure 4b, we show the same genealogical tree, but indicate with different colors the values of the 5-days temperature anomaly in each segment. These values are computed using equation \ref{definition_A} with $T$=5 days and $t_I$ corresponding to the date of each resampling event. In particular red segments indicate a 5-day anomaly above 4.5$^o$K. The chosen threshold of 4.5$^o$K is the median of the distribution of the 5-day temperature anomalies rare event algorithm experiments for Scandinavia (shown in the SI). This value corresponds to the 96.6th percentile of the distribution of the 5-day temperature anomalies in the control run. We can thus consider 5-days periods with temperature anomaly larger than this threshold as heatwave periods, and a succession of 2 or more consecutive heatwave periods as a subseasonal heatwave.

With this definitions, during warmer summers belonging to the second range (red in figure 4a), we see a first subseasonal heatwave that lasts about 20 days in late June-early July, and then a second subseasonal heatwave that last again about 15 days in August. Less extreme summers (black in figure 4a) have instead only one subseasonal heatwave in June-July. The bimodality of the algorithm statistics therefore corresponds to two qualitatively different types of warm summer, with either one or two subseasonal heatwaves. We note that the 2003 warm summer in France was characterized by two separate subseasonal heatwaves in June and August \cite{garcia-herrera_review_2010}.  

\section{Conclusions}\label{conclusions}

We have shown how simulations with the algorithm produce hundreds of times more extremes than a control run for the same computational cost, and allow us to estimate return times of extreme events orders of magnitude larger. This allows to compute precisely statistically significant composite maps of dynamical quantities for warm summers with extremely strong seasonal surface temperature anomalies. In this way we are able to identify rigorously the occurrence of persistent teleconnection patterns of wavenumber 3 during very warm summers. These patterns and the related dynamics can not be properly studied with direct sampling, as the corresponding local anomalies of temperature and geopotential height are statistically significant only over the target region. The improved statistics given by the rare event algorithm instead allows to obtain statistically significant maps and a very clear signal also away from the target region.

Teleconnection patterns similar to what we have obtained here were observed in the first application of the rare event algorithm to a climate model \cite{Ragone&al2018}. In that case the model was an intermediate complexity model, run in perpetual summer setup, and the target area was the whole Europe. In this case we use a much more realistic model with seasonal cycle, calibrated to reproduce present day climate, and we target areas consistent with observed heatwave clusters \cite{Stefanon_2012} and recent cases of very intense heatwaves or warm summer, like the heatwave over France of 2003 and the warm summer over Northern Europe of 2018. It is striking that we obtain similar qualitative results. This supports the idea that these teleconnections at seasonal scale and the corresponding wavenumber 3 dynamics are robust features, mainly dynamical and little affected by the details of the physical parameterizations. 

Comparison with observations is not straightforward, due to the lack of data and the fact that most studies have focused on events at subseasonal scale. Several authors have recently highlighted the role of Rossby waves in determining teleconnection of extreme events \cite{Schubert_2011,LauKim_2012,Petoukhov_2013,Petoukhov_2016}, with particular emphasis on strong heatwaves, e.g. cases over the central USA \cite{Branstator_2013}, Alberta \cite{Petoukhov_2018} and Western Europe \cite{Kornhuber_2019}. These studies typically find patterns with wavenumber 5 to 7 and analyse subseasonal temperature fluctuations. Several different detection methods are used in the literature to identify the dynamics leading to heatwaves, including  projections of heatwave states on typical variability patterns obtained with cluster analysis \cite{Cassou2005} or empirical orthogonal functions \cite{Branstator_2013}, spectral analysis \cite{Petoukhov_2018,Kornhuber_2019} and/or indicators based on resonance models \cite{Petoukhov_2013,Petoukhov_2016}. Merging these approaches with rare event simulations should lead to very promising future studies, aimed at investigating the relation between the seasonal scale structures we find here and the possible role of higher wavenumber stationary Rossby waves proposed in the literature, as well as the possible multimodality in the way of occurrence of some extreme events.

\acknowledgments
This work was granted access to the HPC resources of CINES under the allocation 2019-A0070110575 made by GENCI. The computation of this work were partially performed on the PSMN platform of ENS de Lyon. This work has received funding through the ACADEMICS grant of the IDEXLYON, project of the Université de Lyon, PIA operated by ANR-16-IDEX-0005. The research leading to these results has received funding from the European Research Council under the European Union’s seventh Framework Programme (FP7/2007-2013 Grant Agreement No. 616811). The data related to the figures presented in this paper have been uploaded with the submission for review purposes. We are discussing with CINES the long term storage and open access of the full dataset of our simulations according to the Enabling FAIR data Project guidelines.

\newpage
\clearpage

\begin{figure}\label{statistics}
\includegraphics[width=1\textwidth]{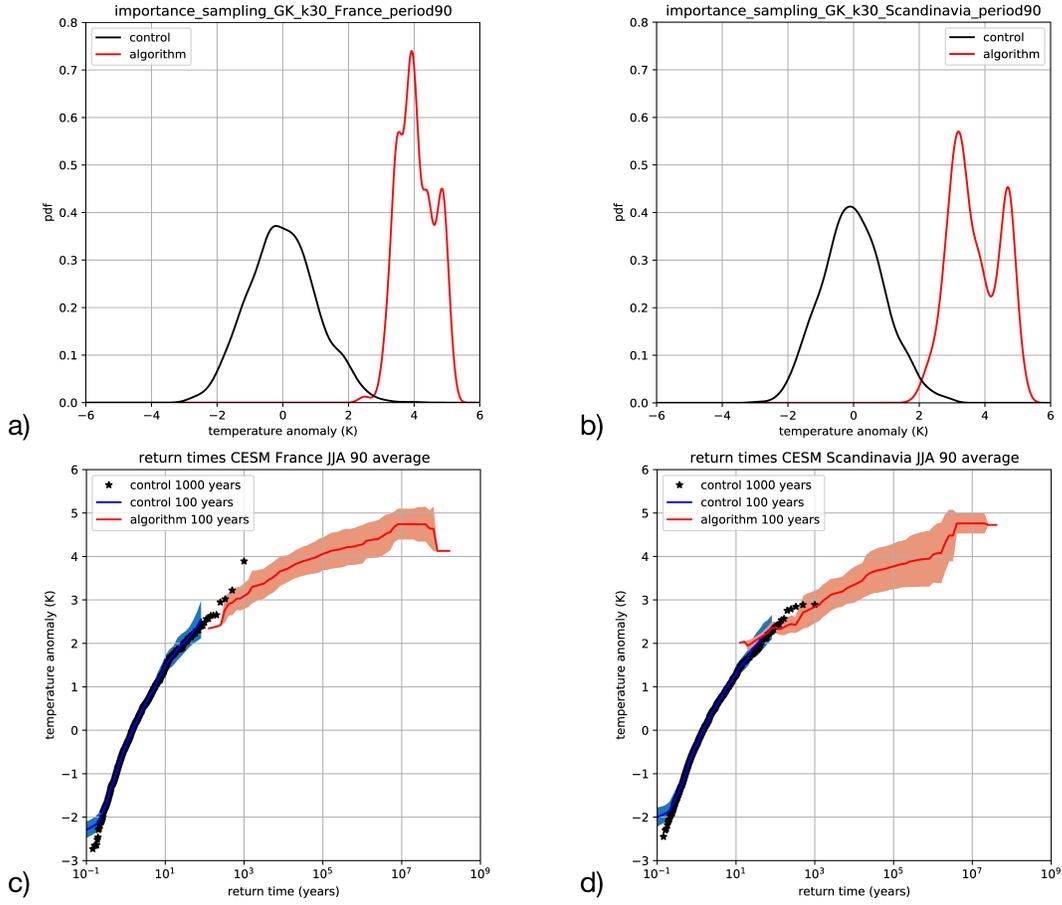} 
\caption{Distribution of seasonal JJA temperatures anomalies averaged over France (a) and Scandinavia (b) from the control run (black) and the rare event algorithm (red). Return times for the seasonal temperature anomalies for France (c) and Scandinavia (d), from the control run (black) and the rare event algorithm (red). The shaded areas, light blue for the control run and light red for the algorithm, correspond to 1 standard deviation of the sample used to compute the estimate (see SI).}
\end{figure}

\newpage
\begin{figure}
\label{composite_JJA_France}
\includegraphics[width=1\textwidth]{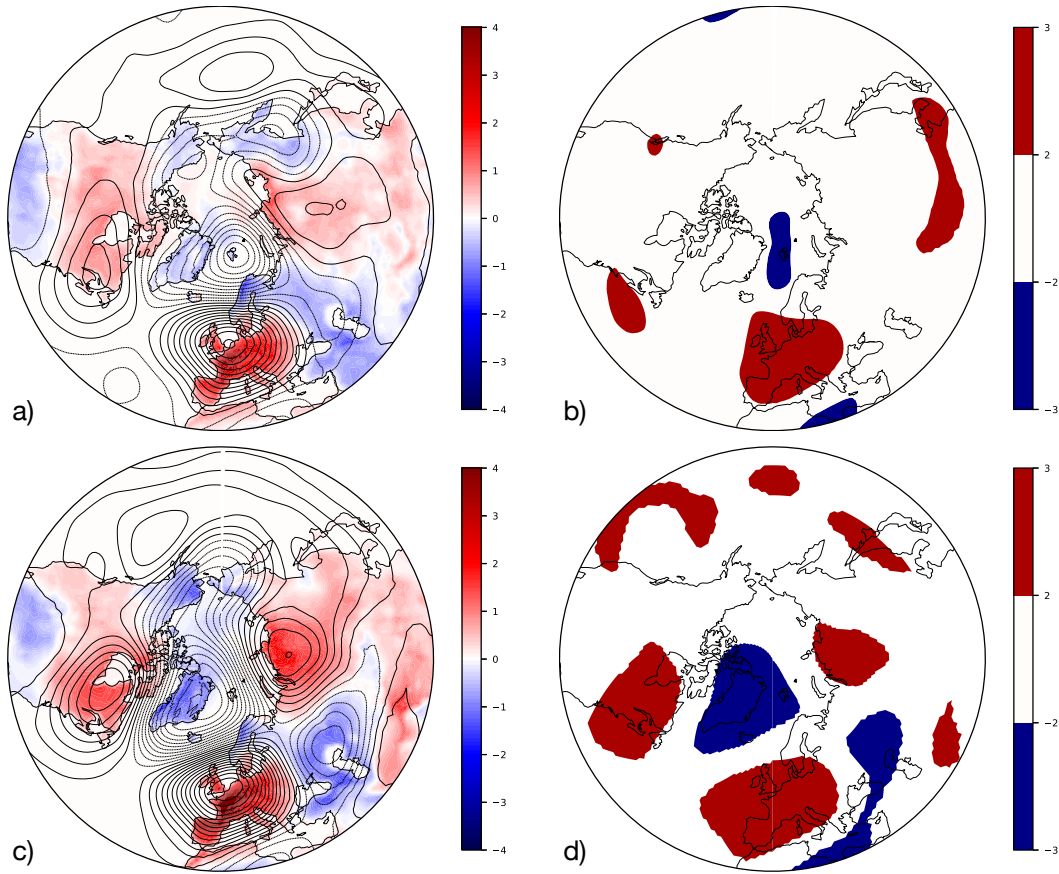} 
\caption{Northern hemisphere composite maps (conditional statistics) for the JJA anomalies of the surface temperature (colors) and 500 hPa geopotential height (contours) for 100-year warm summers over France, in the control run (a) and in the rare event algorithm statistics (c). Panels b) and d) show for the control run and the rare event algorithm respectively the corresponding maps of the $t$ values of the 500 hPa geopotential height anomalies (see the SI).}
\end{figure}

\newpage
\begin{figure}
\includegraphics[width=1\textwidth]{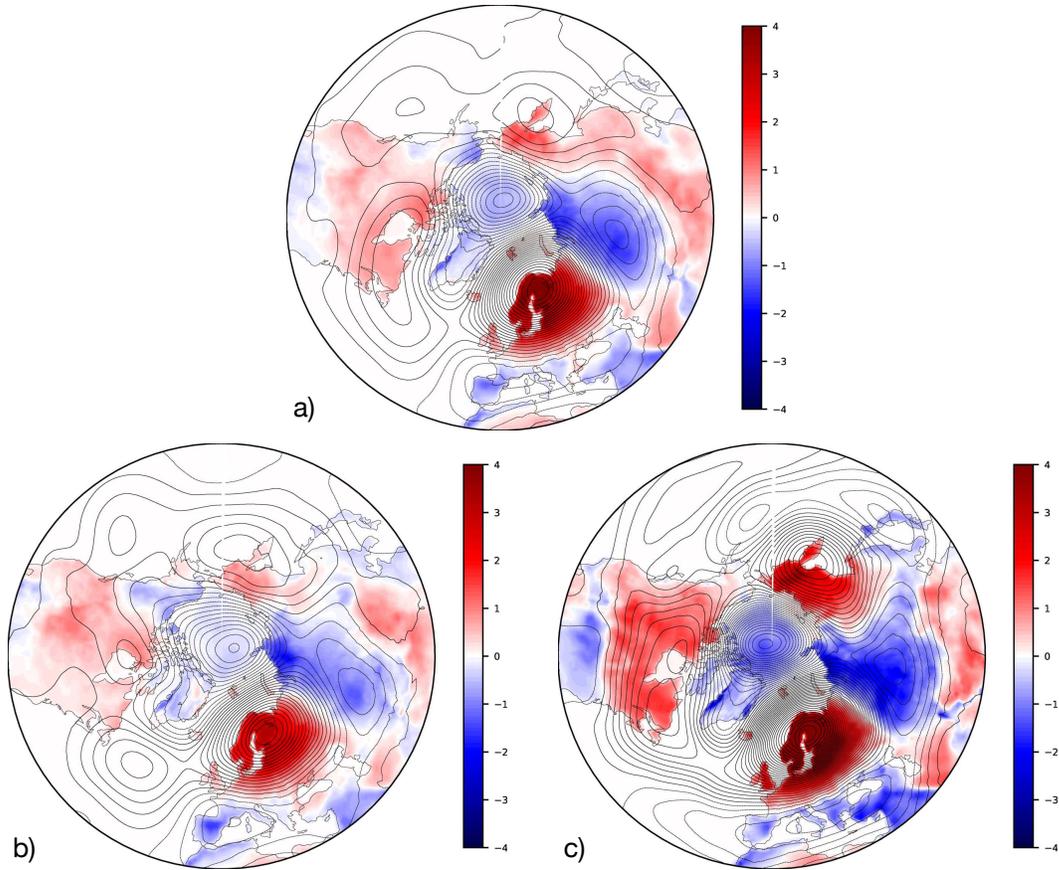} 
\caption{Composite maps for the anomalies of the surface temperature (colors) and 500 hPa geopotential height (contours) for warm summers over Scandinavia with return times larger than 1000 years (a). Panel b) and c) show respectively composite maps for the same variables with the condition of JJA temperature anomaly smaller or higher that 4.2$^o$K.} 
\end{figure}

\newpage
\begin{figure}
\label{trees}
\includegraphics[width=1\textwidth,height=1\textwidth]{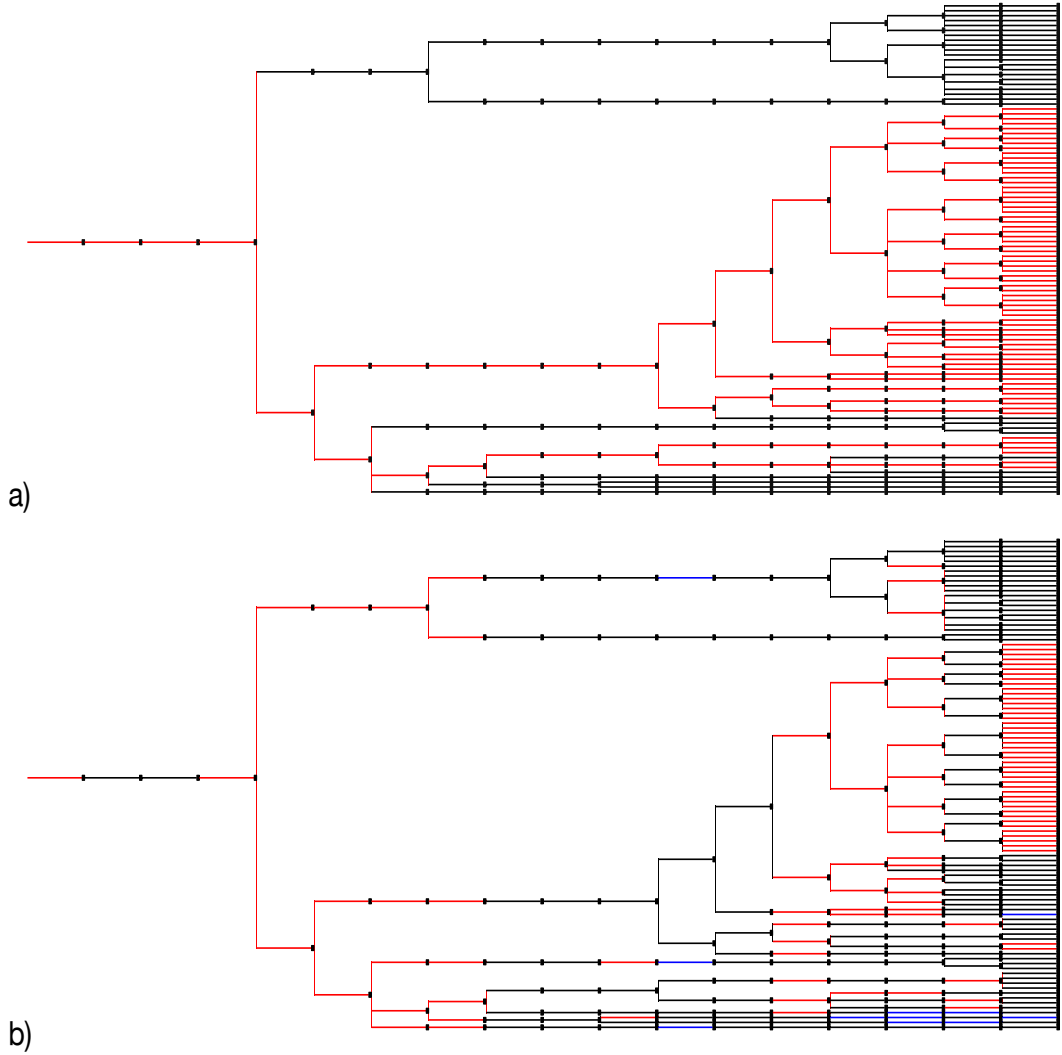} 
\caption{Panel a) shows the genealogical tree of a rare event algorithm experiment, with $N=100$ ensemble members, for selection of Scandinavia seasonal heatwaves. Each broken line, composed by one trunk and its branches up to the last leave, represents a trajectory. The horizontal axis represents time in blocks of $\tau=$ 5 days. The genealogical trees have been created with the help of the ETE Python toolkit \cite{ETE3}. Black lines represent trajectories with $a_{\mbox{\tiny \rm JJA}}<4.2^o$K, red lines trajectories with $a_{\mbox{\tiny \rm JJA}}>4.2^o$K. Panel b): the same, but with different colors for each 5 day period: blue corresponds to negative 5-day time averaged anomalies, black between 0 and 4.5$^o$K, and red above 4.5$^o$K.}
\end{figure}

\newpage
\clearpage

\bibliography{ref,references}

\begin{thebibliography}{}

\bibitem [\protect \citeauthoryear {%
Agha~Kouchak%
}{%
Agha~Kouchak%
}{%
{\protect \APACyear {2012}}%
}]{%
AghaKouchak2012}
\APACinsertmetastar {%
AghaKouchak2012}%
\begin{APACrefauthors}%
Agha~Kouchak, A.%
\end{APACrefauthors}%
\unskip\
\newblock
\APACrefYear{2012}.
\newblock
\APACrefbtitle {Extremes in a changing climate detection, analysis and
  uncertainty} {Extremes in a changing climate detection, analysis and
  uncertainty}.
\newblock
\APACaddressPublisher{Dordrecht; New York}{Springer}.
\PrintBackRefs{\CurrentBib}

\bibitem [\protect \citeauthoryear {%
Barriopedro%
, Fischer%
, Luterbacher%
, Trigo%
\BCBL {}\ \BBA {} Garcia-Herrera%
}{%
Barriopedro%
\ \protect \BOthers {.}}{%
{\protect \APACyear {2011}}%
}]{%
Barriopedro_2011}
\APACinsertmetastar {%
Barriopedro_2011}%
\begin{APACrefauthors}%
Barriopedro, D.%
, Fischer, E.%
, Luterbacher, J.%
, Trigo, R.%
\BCBL {}\ \BBA {} Garcia-Herrera, R.%
\end{APACrefauthors}%
\unskip\
\newblock
\APACrefYearMonthDay{2011}{}{}.
\newblock
{\BBOQ}\APACrefatitle {Redrawing the temperature record map of Europe}
  {Redrawing the temperature record map of europe}.{\BBCQ}
\newblock
\APACjournalVolNumPages{Science}{332}{}{220–224}.
\PrintBackRefs{\CurrentBib}

\bibitem [\protect \citeauthoryear {%
Boers%
, Goswami%
\BCBL {}\ \BBA {} Rheinwalt%
}{%
Boers%
\ \protect \BOthers {.}}{%
{\protect \APACyear {2019}}%
}]{%
Boers2019}
\APACinsertmetastar {%
Boers2019}%
\begin{APACrefauthors}%
Boers, N.%
, Goswami, B.%
\BCBL {}\ \BBA {} Rheinwalt, A\BPBI e\BPBI a.%
\end{APACrefauthors}%
\unskip\
\newblock
\APACrefYearMonthDay{2019}{}{}.
\newblock
{\BBOQ}\APACrefatitle {Complex networks reveal global pattern of
  extreme-rainfall teleconnections} {Complex networks reveal global pattern of
  extreme-rainfall teleconnections}.{\BBCQ}
\newblock
\APACjournalVolNumPages{Nature}{566}{}{373-377}.
\newblock
\begin{APACrefDOI} \doi{10.1038/s41586-018-0872-x} \end{APACrefDOI}
\PrintBackRefs{\CurrentBib}

\bibitem [\protect \citeauthoryear {%
Bouchet%
, Rolland%
\BCBL {}\ \BBA {} Simonnet%
}{%
Bouchet%
\ \protect \BOthers {.}}{%
{\protect \APACyear {2019}}%
}]{%
Bouchet_Rolland_Simonnet_2019:C}
\APACinsertmetastar {%
Bouchet_Rolland_Simonnet_2019:C}%
\begin{APACrefauthors}%
Bouchet, F.%
, Rolland, J.%
\BCBL {}\ \BBA {} Simonnet, E.%
\end{APACrefauthors}%
\unskip\
\newblock
\APACrefYearMonthDay{2019}{}{}.
\newblock
{\BBOQ}\APACrefatitle {Rare Event Algorithm Links Transitions in Turbulent
  Flows with Activated Nucleations} {Rare event algorithm links transitions in
  turbulent flows with activated nucleations}.{\BBCQ}
\newblock
\APACjournalVolNumPages{Physical Review Letters}{122}{7}{074502}.
\PrintBackRefs{\CurrentBib}

\bibitem [\protect \citeauthoryear {%
Camargo%
\ \BBA {} Seth%
}{%
Camargo%
\ \BBA {} Seth%
}{%
{\protect \APACyear {2016}}%
}]{%
Camargo2016}
\APACinsertmetastar {%
Camargo2016}%
\begin{APACrefauthors}%
Camargo, S\BPBI J.%
\BCBT {}\ \BBA {} Seth, A.%
\end{APACrefauthors}%
\unskip\
\newblock
\APACrefYearMonthDay{2016}{}{}.
\newblock
{\BBOQ}\APACrefatitle {Hottest summers the new normal} {Hottest summers the new
  normal}.{\BBCQ}
\newblock
\APACjournalVolNumPages{Environ. Res. Lett.}{11}{}{081001}.
\PrintBackRefs{\CurrentBib}

\bibitem [\protect \citeauthoryear {%
Cassou%
, Terray%
\BCBL {}\ \BBA {} Phillips%
}{%
Cassou%
\ \protect \BOthers {.}}{%
{\protect \APACyear {2005}}%
}]{%
Cassou2005}
\APACinsertmetastar {%
Cassou2005}%
\begin{APACrefauthors}%
Cassou, C.%
, Terray, L.%
\BCBL {}\ \BBA {} Phillips, A\BPBI S.%
\end{APACrefauthors}%
\unskip\
\newblock
\APACrefYearMonthDay{2005}{}{}.
\newblock
{\BBOQ}\APACrefatitle {Tropical Atlantic Influence on European Heat Waves}
  {Tropical atlantic influence on european heat waves}.{\BBCQ}
\newblock
\APACjournalVolNumPages{Journal of Climate}{18}{15}{2805-2811}.
\newblock
\begin{APACrefDOI} \doi{10.1175/JCLI3506.1} \end{APACrefDOI}
\PrintBackRefs{\CurrentBib}

\bibitem [\protect \citeauthoryear {%
Coumou%
\ \BBA {} Rahmstorf%
}{%
Coumou%
\ \BBA {} Rahmstorf%
}{%
{\protect \APACyear {2012}}%
}]{%
coumou_decade_2012}
\APACinsertmetastar {%
coumou_decade_2012}%
\begin{APACrefauthors}%
Coumou, D.%
\BCBT {}\ \BBA {} Rahmstorf, S.%
\end{APACrefauthors}%
\unskip\
\newblock
\APACrefYearMonthDay{2012}{}{}.
\newblock
{\BBOQ}\APACrefatitle {A decade of weather extremes} {A decade of weather
  extremes}.{\BBCQ}
\newblock
\APACjournalVolNumPages{Nature Climate Change}{2}{7}{491-496}.
\newblock
\begin{APACrefDOI} \doi{10.1038/nclimate1452} \end{APACrefDOI}
\PrintBackRefs{\CurrentBib}

\bibitem [\protect \citeauthoryear {%
D'Andrea%
, Drobinski%
\BCBL {}\ \BBA {} Stefanon%
}{%
D'Andrea%
\ \protect \BOthers {.}}{%
{\protect \APACyear {2016}}%
}]{%
DAndrea&al2016}
\APACinsertmetastar {%
DAndrea&al2016}%
\begin{APACrefauthors}%
D'Andrea, F.%
, Drobinski, P.%
\BCBL {}\ \BBA {} Stefanon, M.%
\end{APACrefauthors}%
\unskip\
\newblock
\APACrefYearMonthDay{2016}{}{}.
\newblock
{\BBOQ}\APACrefatitle {European heat waves: the effect of soil moisture,
  vegetation, and land use} {European heat waves: the effect of soil moisture,
  vegetation, and land use}.{\BBCQ}
\newblock
\BIn{} \APACrefbtitle {Dynamics and Predictability of Large-Scale, High-Impact
  Weather and Climate Events} {Dynamics and predictability of large-scale,
  high-impact weather and climate events}\ (\BPG~185-197).
\newblock
\APACaddressPublisher{}{Cambridge University Press}.
\PrintBackRefs{\CurrentBib}

\bibitem [\protect \citeauthoryear {%
Della-Marta%
\ \protect \BOthers {.}}{%
Della-Marta%
\ \protect \BOthers {.}}{%
{\protect \APACyear {2007}}%
}]{%
della-marta_summer_2007}
\APACinsertmetastar {%
della-marta_summer_2007}%
\begin{APACrefauthors}%
Della-Marta, P\BPBI M.%
, Luterbacher, J.%
, {von Weissenfluh}, H.%
, Xoplaki, E.%
, Brunet, M.%
\BCBL {}\ \BBA {} Wanner, H.%
\end{APACrefauthors}%
\unskip\
\newblock
\APACrefYearMonthDay{2007}{}{}.
\newblock
{\BBOQ}\APACrefatitle {Summer heat waves over western {{Europe}}
  1880{\textendash}2003, their relationship to large-scale forcings and
  predictability} {Summer heat waves over western {{Europe}}
  1880{\textendash}2003, their relationship to large-scale forcings and
  predictability}.{\BBCQ}
\newblock
\APACjournalVolNumPages{Climate Dynamics}{29}{2-3}{251-275}.
\newblock
\begin{APACrefDOI} \doi{10.1007/s00382-007-0233-1} \end{APACrefDOI}
\PrintBackRefs{\CurrentBib}

\bibitem [\protect \citeauthoryear {%
Del~Moral%
, Garnier%
\BCBL {}\ \protect \BOthers {.}}{%
Del~Moral%
\ \protect \BOthers {.}}{%
{\protect \APACyear {2005}}%
}]{%
del2005genealogical}
\APACinsertmetastar {%
del2005genealogical}%
\begin{APACrefauthors}%
Del~Moral, P.%
, Garnier, J.%
\BCBL {}\ \BOthersPeriod {.}\end{APACrefauthors}%
\unskip\
\newblock
\APACrefYearMonthDay{2005}{}{}.
\newblock
{\BBOQ}\APACrefatitle {Genealogical particle analysis of rare events}
  {Genealogical particle analysis of rare events}.{\BBCQ}
\newblock
\APACjournalVolNumPages{The Annals of Applied Probability}{15}{4}{2496--2534}.
\PrintBackRefs{\CurrentBib}

\bibitem [\protect \citeauthoryear {%
Dosio%
, Mentaschi%
, Fischer%
\BCBL {}\ \BBA {} Wyser%
}{%
Dosio%
\ \protect \BOthers {.}}{%
{\protect \APACyear {2018}}%
}]{%
Dosio_et_al_2018}
\APACinsertmetastar {%
Dosio_et_al_2018}%
\begin{APACrefauthors}%
Dosio, A.%
, Mentaschi, L.%
, Fischer, E\BPBI M.%
\BCBL {}\ \BBA {} Wyser, K.%
\end{APACrefauthors}%
\unskip\
\newblock
\APACrefYearMonthDay{2018}{}{}.
\newblock
{\BBOQ}\APACrefatitle {Extreme heat waves under 1.5 C and 2 C global warming}
  {Extreme heat waves under 1.5 c and 2 c global warming}.{\BBCQ}
\newblock
\APACjournalVolNumPages{Environ. Res. Lett.}{13}{}{054006}.
\newblock
\begin{APACrefDOI} \doi{10.1088/1748-9326/aab827} \end{APACrefDOI}
\PrintBackRefs{\CurrentBib}

\bibitem [\protect \citeauthoryear {%
Ebener%
, Margazoglou%
, Friedrich%
, Biferale%
\BCBL {}\ \BBA {} Grauer%
}{%
Ebener%
\ \protect \BOthers {.}}{%
{\protect \APACyear {2019}}%
}]{%
Grauer}
\APACinsertmetastar {%
Grauer}%
\begin{APACrefauthors}%
Ebener, L.%
, Margazoglou, G.%
, Friedrich, J.%
, Biferale, L.%
\BCBL {}\ \BBA {} Grauer, R.%
\end{APACrefauthors}%
\unskip\
\newblock
\APACrefYearMonthDay{2019}{}{}.
\newblock
{\BBOQ}\APACrefatitle {Instanton base importance sampling for rare events in
  stochastic PDEs} {Instanton base importance sampling for rare events in
  stochastic pdes}.{\BBCQ}
\newblock
\APACjournalVolNumPages{Chaos: An Interdisciplinary Journal of Nonlinear
  Science}{}{}{}.
\PrintBackRefs{\CurrentBib}

\bibitem [\protect \citeauthoryear {%
Garc{\'\i}a-Herrera%
, D{\'\i}az%
, Trigo%
, Luterbacher%
\BCBL {}\ \BBA {} Fischer%
}{%
Garc{\'\i}a-Herrera%
\ \protect \BOthers {.}}{%
{\protect \APACyear {2010}}%
}]{%
garcia-herrera_review_2010}
\APACinsertmetastar {%
garcia-herrera_review_2010}%
\begin{APACrefauthors}%
Garc{\'\i}a-Herrera, R.%
, D{\'\i}az, J.%
, Trigo, R\BPBI M.%
, Luterbacher, J.%
\BCBL {}\ \BBA {} Fischer, E\BPBI M.%
\end{APACrefauthors}%
\unskip\
\newblock
\APACrefYearMonthDay{2010}{}{}.
\newblock
{\BBOQ}\APACrefatitle {A {{Review}} of the {{European Summer Heat Wave}} of
  2003} {A {{Review}} of the {{European Summer Heat Wave}} of 2003}.{\BBCQ}
\newblock
\APACjournalVolNumPages{Critical Reviews in Environmental Science and
  Technology}{40}{4}{267-306}.
\newblock
\begin{APACrefDOI} \doi{10.1080/10643380802238137} \end{APACrefDOI}
\PrintBackRefs{\CurrentBib}

\bibitem [\protect \citeauthoryear {%
Giardina%
, Kurchan%
, Lecomte%
\BCBL {}\ \BBA {} Tailleur%
}{%
Giardina%
\ \protect \BOthers {.}}{%
{\protect \APACyear {2011}}%
}]{%
giardina_simulating_2011}
\APACinsertmetastar {%
giardina_simulating_2011}%
\begin{APACrefauthors}%
Giardina, C.%
, Kurchan, J.%
, Lecomte, V.%
\BCBL {}\ \BBA {} Tailleur, J.%
\end{APACrefauthors}%
\unskip\
\newblock
\APACrefYearMonthDay{2011}{}{}.
\newblock
{\BBOQ}\APACrefatitle {Simulating {{Rare Events}} in {{Dynamical Processes}}}
  {Simulating {{Rare Events}} in {{Dynamical Processes}}}.{\BBCQ}
\newblock
\APACjournalVolNumPages{Journal of Statistical Physics}{145}{4}{787-811}.
\newblock
\begin{APACrefDOI} \doi{10.1007/s10955-011-0350-4} \end{APACrefDOI}
\PrintBackRefs{\CurrentBib}

\bibitem [\protect \citeauthoryear {%
Grafke%
, Grauer%
\BCBL {}\ \BBA {} Sch\"afer%
}{%
Grafke%
\ \protect \BOthers {.}}{%
{\protect \APACyear {2015}}%
}]{%
Grafke}
\APACinsertmetastar {%
Grafke}%
\begin{APACrefauthors}%
Grafke, T.%
, Grauer, R.%
\BCBL {}\ \BBA {} Sch\"afer, T.%
\end{APACrefauthors}%
\unskip\
\newblock
\APACrefYearMonthDay{2015}{}{}.
\newblock
{\BBOQ}\APACrefatitle {The instanton method and its numerical implementation in
  fluid mechanics} {The instanton method and its numerical implementation in
  fluid mechanics}.{\BBCQ}
\newblock
\APACjournalVolNumPages{J.Phys.A:Math.Theor.}{48}{}{333001}.
\PrintBackRefs{\CurrentBib}

\bibitem [\protect \citeauthoryear {%
Horton%
, Mankin%
\BCBL {}\ \BBA {} Lesk%
}{%
Horton%
\ \protect \BOthers {.}}{%
{\protect \APACyear {2016}}%
}]{%
Horton2016}
\APACinsertmetastar {%
Horton2016}%
\begin{APACrefauthors}%
Horton, R.%
, Mankin, J.%
\BCBL {}\ \BBA {} Lesk, C\BPBI e\BPBI a.%
\end{APACrefauthors}%
\unskip\
\newblock
\APACrefYearMonthDay{2016}{}{}.
\newblock
{\BBOQ}\APACrefatitle {A Review of Recent Advances in Research on Extreme Heat
  Events} {A review of recent advances in research on extreme heat
  events}.{\BBCQ}
\newblock
\APACjournalVolNumPages{Curr Clim Change Rep}{2}{}{242–259}.
\newblock
\begin{APACrefDOI} \doi{10.1007/s40641-016-0042-x} \end{APACrefDOI}
\PrintBackRefs{\CurrentBib}

\bibitem [\protect \citeauthoryear {%
Hoskins%
\ \BBA {} Woollings%
}{%
Hoskins%
\ \BBA {} Woollings%
}{%
{\protect \APACyear {2015}}%
}]{%
Hoskins2015}
\APACinsertmetastar {%
Hoskins2015}%
\begin{APACrefauthors}%
Hoskins, B.%
\BCBT {}\ \BBA {} Woollings, T.%
\end{APACrefauthors}%
\unskip\
\newblock
\APACrefYearMonthDay{2015}{}{}.
\newblock
{\BBOQ}\APACrefatitle {Persistent Extratropical Regimes and Climate Extremes}
  {Persistent extratropical regimes and climate extremes}.{\BBCQ}
\newblock
\APACjournalVolNumPages{Current Climate Change Reports}{1}{3}{115--124}.
\newblock
\begin{APACrefDOI} \doi{10.1007/s40641-015-0020-8} \end{APACrefDOI}
\PrintBackRefs{\CurrentBib}

\bibitem [\protect \citeauthoryear {%
Huerta-Cepas%
, Serra%
\BCBL {}\ \BBA {} Bork%
}{%
Huerta-Cepas%
\ \protect \BOthers {.}}{%
{\protect \APACyear {2016}}%
}]{%
ETE3}
\APACinsertmetastar {%
ETE3}%
\begin{APACrefauthors}%
Huerta-Cepas, J.%
, Serra, F.%
\BCBL {}\ \BBA {} Bork, P.%
\end{APACrefauthors}%
\unskip\
\newblock
\APACrefYearMonthDay{2016}{}{}.
\newblock
{\BBOQ}\APACrefatitle {ETE 3: Reconstruction, Analysis, and Visualization of
  Phylogenomic Data} {Ete 3: Reconstruction, analysis, and visualization of
  phylogenomic data}.{\BBCQ}
\newblock
\APACjournalVolNumPages{Molecular Biology and Evolution}{33}{6}{1635–1638}.
\newblock
\begin{APACrefDOI} \doi{10.1093/molbev/msw046} \end{APACrefDOI}
\PrintBackRefs{\CurrentBib}

\bibitem [\protect \citeauthoryear {%
Hurrell%
\ \protect \BOthers {.}}{%
Hurrell%
\ \protect \BOthers {.}}{%
{\protect \APACyear {2013}}%
}]{%
CESM}
\APACinsertmetastar {%
CESM}%
\begin{APACrefauthors}%
Hurrell, J\BPBI W.%
, Holland, M\BPBI M.%
, Gent, P\BPBI R.%
, Ghan, S.%
, Kay, J\BPBI E.%
, Kushner, P\BPBI J.%
\BDBL {}Marshall, S.%
\end{APACrefauthors}%
\unskip\
\newblock
\APACrefYearMonthDay{2013}{}{}.
\newblock
{\BBOQ}\APACrefatitle {The Community Earth System Model: A Framework for
  Collaborative Research} {The community earth system model: A framework for
  collaborative research}.{\BBCQ}
\newblock
\APACjournalVolNumPages{Bulletin of the American Meteorological
  Society}{94}{9}{1339-1360}.
\newblock
\begin{APACrefDOI} \doi{10.1175/BAMS-D-12-00121.1} \end{APACrefDOI}
\PrintBackRefs{\CurrentBib}

\bibitem [\protect \citeauthoryear {%
IPCC%
}{%
IPCC%
}{%
{\protect \APACyear {2012}}%
}]{%
IPCC_2012}
\APACinsertmetastar {%
IPCC_2012}%
\begin{APACrefauthors}%
IPCC.%
\end{APACrefauthors}%
\unskip\
\newblock
\APACrefYear{2012}.
\newblock
\APACrefbtitle {Managing the risks of extreme events and disasters to advance
  climate change adaption: special report of the {{Intergovernmental Panel}} on
  {{Climate Change}}} {Managing the risks of extreme events and disasters to
  advance climate change adaption: special report of the {{Intergovernmental
  Panel}} on {{Climate Change}}}.
\newblock
\APACaddressPublisher{New York, NY}{{Cambridge University Press}}.
\PrintBackRefs{\CurrentBib}

\bibitem [\protect \citeauthoryear {%
IPCC%
}{%
IPCC%
}{%
{\protect \APACyear {2013}}%
}]{%
IPCC_2013}
\APACinsertmetastar {%
IPCC_2013}%
\begin{APACrefauthors}%
IPCC.%
\end{APACrefauthors}%
\unskip\
\newblock
\APACrefYear{2013}.
\newblock
\APACrefbtitle {Climate {{Change}} 2013: {{The Physical Science Basis}}.
  {{Contribution}} of {{Working Group I}} to the {{Fifth Assessment Report}} of
  the {{Intergovernmental Panel}} on {{Climate Change}}} {Climate {{Change}}
  2013: {{The Physical Science Basis}}. {{Contribution}} of {{Working Group I}}
  to the {{Fifth Assessment Report}} of the {{Intergovernmental Panel}} on
  {{Climate Change}}}.
\newblock
\APACaddressPublisher{Cambridge, United Kingdom and New York, NY,
  USA}{{Cambridge University Press}}.
\PrintBackRefs{\CurrentBib}

\bibitem [\protect \citeauthoryear {%
Jezequel%
, Yiou%
\BCBL {}\ \BBA {} Radanovics%
}{%
Jezequel%
\ \protect \BOthers {.}}{%
{\protect \APACyear {2018}}%
}]{%
Jezequel_et_al_2018}
\APACinsertmetastar {%
Jezequel_et_al_2018}%
\begin{APACrefauthors}%
Jezequel, A.%
, Yiou, P.%
\BCBL {}\ \BBA {} Radanovics, S.%
\end{APACrefauthors}%
\unskip\
\newblock
\APACrefYearMonthDay{2018}{}{}.
\newblock
{\BBOQ}\APACrefatitle {Role of circulation in European heatwaves using flow
  analogues} {Role of circulation in european heatwaves using flow
  analogues}.{\BBCQ}
\newblock
\APACjournalVolNumPages{Clim Dyn}{50}{}{1145–1159}.
\newblock
\begin{APACrefDOI} \doi{10.1007/s00382-017-3667-0} \end{APACrefDOI}
\PrintBackRefs{\CurrentBib}

\bibitem [\protect \citeauthoryear {%
Kornhuber%
\ \protect \BOthers {.}}{%
Kornhuber%
\ \protect \BOthers {.}}{%
{\protect \APACyear {2019}}%
}]{%
Kornhuber_2019}
\APACinsertmetastar {%
Kornhuber_2019}%
\begin{APACrefauthors}%
Kornhuber, K.%
, Osprey, S.%
, Coumou, D.%
, Petri, S.%
, Petoukhov, V.%
, Rahmstorf, S.%
\BCBL {}\ \BBA {} Gray, L.%
\end{APACrefauthors}%
\unskip\
\newblock
\APACrefYearMonthDay{2019}{}{}.
\newblock
{\BBOQ}\APACrefatitle {Extreme weather events in early summer 2018 connected by
  a recurrent hemispheric wave-7 pattern} {Extreme weather events in early
  summer 2018 connected by a recurrent hemispheric wave-7 pattern}.{\BBCQ}
\newblock
\APACjournalVolNumPages{Environmental Research Letters}{14}{5}{054002}.
\newblock
\begin{APACrefDOI} \doi{10.1088/1748-9326/ab13bf} \end{APACrefDOI}
\PrintBackRefs{\CurrentBib}

\bibitem [\protect \citeauthoryear {%
Lau%
\ \BBA {} Kim%
}{%
Lau%
\ \BBA {} Kim%
}{%
{\protect \APACyear {2012}}%
}]{%
LauKim_2012}
\APACinsertmetastar {%
LauKim_2012}%
\begin{APACrefauthors}%
Lau, W.%
\BCBT {}\ \BBA {} Kim, K\BHBI M.%
\end{APACrefauthors}%
\unskip\
\newblock
\APACrefYearMonthDay{2012}{02}{}.
\newblock
{\BBOQ}\APACrefatitle {The 2010 Pakistan Flood and Russian Heat Wave:
  Teleconnection of Hydrometeorological Extremes} {The 2010 pakistan flood and
  russian heat wave: Teleconnection of hydrometeorological extremes}.{\BBCQ}
\newblock
\APACjournalVolNumPages{Journal of Hydrometeorology}{13}{}{392-403}.
\newblock
\begin{APACrefDOI} \doi{10.1175/JHM-D-11-016.1} \end{APACrefDOI}
\PrintBackRefs{\CurrentBib}

\bibitem [\protect \citeauthoryear {%
Laurie%
\ \BBA {} Bouchet%
}{%
Laurie%
\ \BBA {} Bouchet%
}{%
{\protect \APACyear {2015}}%
}]{%
Laurie}
\APACinsertmetastar {%
Laurie}%
\begin{APACrefauthors}%
Laurie, J.%
\BCBT {}\ \BBA {} Bouchet, F.%
\end{APACrefauthors}%
\unskip\
\newblock
\APACrefYearMonthDay{2015}{}{}.
\newblock
{\BBOQ}\APACrefatitle {Computation of rare transitions in the barotropic
  quasi-geostrophic equations} {Computation of rare transitions in the
  barotropic quasi-geostrophic equations}.{\BBCQ}
\newblock
\APACjournalVolNumPages{New J. of Phys.}{17}{}{015009}.
\PrintBackRefs{\CurrentBib}

\bibitem [\protect \citeauthoryear {%
Lestang%
, Bouchet%
\BCBL {}\ \BBA {} L{\'e}v{\^e}que%
}{%
Lestang%
\ \protect \BOthers {.}}{%
{\protect \APACyear {2020}}%
}]{%
lestang2020numerical}
\APACinsertmetastar {%
lestang2020numerical}%
\begin{APACrefauthors}%
Lestang, T.%
, Bouchet, F.%
\BCBL {}\ \BBA {} L{\'e}v{\^e}que, E.%
\end{APACrefauthors}%
\unskip\
\newblock
\APACrefYearMonthDay{2020}{}{}.
\newblock
{\BBOQ}\APACrefatitle {Numerical study of extreme mechanical force exerted by a
  turbulent flow on a bluff body by direct and rare-event sampling techniques}
  {Numerical study of extreme mechanical force exerted by a turbulent flow on a
  bluff body by direct and rare-event sampling techniques}.{\BBCQ}
\newblock
\APACjournalVolNumPages{Journal of Fluid Mechanics}{895}{}{}.
\PrintBackRefs{\CurrentBib}

\bibitem [\protect \citeauthoryear {%
Lestang%
, Ragone%
, Br{\'{e}}hier%
, Herbert%
\BCBL {}\ \BBA {} Bouchet%
}{%
Lestang%
\ \protect \BOthers {.}}{%
{\protect \APACyear {2018}}%
}]{%
Lestang_2018}
\APACinsertmetastar {%
Lestang_2018}%
\begin{APACrefauthors}%
Lestang, T.%
, Ragone, F.%
, Br{\'{e}}hier, C\BHBI E.%
, Herbert, C.%
\BCBL {}\ \BBA {} Bouchet, F.%
\end{APACrefauthors}%
\unskip\
\newblock
\APACrefYearMonthDay{2018}{}{}.
\newblock
{\BBOQ}\APACrefatitle {Computing return times or return periods with rare event
  algorithms} {Computing return times or return periods with rare event
  algorithms}.{\BBCQ}
\newblock
\APACjournalVolNumPages{Journal of Statistical Mechanics: Theory and
  Experiment}{2018}{4}{043213}.
\newblock
\begin{APACrefDOI} \doi{10.1088/1742-5468/aab856} \end{APACrefDOI}
\PrintBackRefs{\CurrentBib}

\bibitem [\protect \citeauthoryear {%
Luterbacher%
, Dietrich%
, Xoplaki%
, Grosjean%
\BCBL {}\ \BBA {} Wanner%
}{%
Luterbacher%
\ \protect \BOthers {.}}{%
{\protect \APACyear {2004}}%
}]{%
Luterbacher2004}
\APACinsertmetastar {%
Luterbacher2004}%
\begin{APACrefauthors}%
Luterbacher, J.%
, Dietrich, D.%
, Xoplaki, E.%
, Grosjean, M.%
\BCBL {}\ \BBA {} Wanner, H.%
\end{APACrefauthors}%
\unskip\
\newblock
\APACrefYearMonthDay{2004}{}{}.
\newblock
{\BBOQ}\APACrefatitle {European Seasonal and Annual Temperature Variability,
  Trends, and Extremes Since 1500} {European seasonal and annual temperature
  variability, trends, and extremes since 1500}.{\BBCQ}
\newblock
\APACjournalVolNumPages{Science}{303}{5663}{1499-1503}.
\newblock
\begin{APACrefDOI} \doi{10.1126/science.1093877} \end{APACrefDOI}
\PrintBackRefs{\CurrentBib}

\bibitem [\protect \citeauthoryear {%
Mueller%
, Zhang%
\BCBL {}\ \BBA {} Zwiers%
}{%
Mueller%
\ \protect \BOthers {.}}{%
{\protect \APACyear {2016}}%
}]{%
Mueller2016}
\APACinsertmetastar {%
Mueller2016}%
\begin{APACrefauthors}%
Mueller, B.%
, Zhang, X.%
\BCBL {}\ \BBA {} Zwiers, F\BPBI W.%
\end{APACrefauthors}%
\unskip\
\newblock
\APACrefYearMonthDay{2016}{}{}.
\newblock
{\BBOQ}\APACrefatitle {Historically hottest summers projected to be the norm
  for more than half of the world's population within 20 years} {Historically
  hottest summers projected to be the norm for more than half of the world's
  population within 20 years}.{\BBCQ}
\newblock
\APACjournalVolNumPages{Environ. Res. Lett.}{11}{}{044011}.
\PrintBackRefs{\CurrentBib}

\bibitem [\protect \citeauthoryear {%
Otto%
, Massey%
, Van~Oldenborgh%
, Jones%
\BCBL {}\ \BBA {} Allen%
}{%
Otto%
\ \protect \BOthers {.}}{%
{\protect \APACyear {2012}}%
}]{%
Otto_2012}
\APACinsertmetastar {%
Otto_2012}%
\begin{APACrefauthors}%
Otto, F.%
, Massey, N.%
, Van~Oldenborgh, G\BPBI J.%
, Jones, R.%
\BCBL {}\ \BBA {} Allen, M.%
\end{APACrefauthors}%
\unskip\
\newblock
\APACrefYearMonthDay{2012}{02}{}.
\newblock
{\BBOQ}\APACrefatitle {Reconciling two approaches to attribution of the 2010
  Russian heat wave} {Reconciling two approaches to attribution of the 2010
  russian heat wave}.{\BBCQ}
\newblock
\APACjournalVolNumPages{Geophysical Research Letters}{39}{}{4702-}.
\newblock
\begin{APACrefDOI} \doi{10.1029/2011GL050422} \end{APACrefDOI}
\PrintBackRefs{\CurrentBib}

\bibitem [\protect \citeauthoryear {%
{Perkins}%
}{%
{Perkins}%
}{%
{\protect \APACyear {2015}}%
}]{%
Perkins2015}
\APACinsertmetastar {%
Perkins2015}%
\begin{APACrefauthors}%
{Perkins}, S\BPBI E.%
\end{APACrefauthors}%
\unskip\
\newblock
\APACrefYearMonthDay{2015}{}{}.
\newblock
{\BBOQ}\APACrefatitle {{A review on the scientific understanding of
  heatwaves-Their measurement, driving mechanisms, and changes at the global
  scale}} {{A review on the scientific understanding of heatwaves-Their
  measurement, driving mechanisms, and changes at the global scale}}.{\BBCQ}
\newblock
\APACjournalVolNumPages{Atmospheric Research}{164}{}{242-267}.
\newblock
\begin{APACrefDOI} \doi{10.1016/j.atmosres.2015.05.014} \end{APACrefDOI}
\PrintBackRefs{\CurrentBib}

\bibitem [\protect \citeauthoryear {%
Petoukhov%
\ \protect \BOthers {.}}{%
Petoukhov%
\ \protect \BOthers {.}}{%
{\protect \APACyear {2018}}%
}]{%
Petoukhov_2018}
\APACinsertmetastar {%
Petoukhov_2018}%
\begin{APACrefauthors}%
Petoukhov, V.%
, Petri, S.%
, Kornhuber, K.%
, Thonicke, K.%
, Coumou, D.%
\BCBL {}\ \BBA {} Schellnhuber, H.%
\end{APACrefauthors}%
\unskip\
\newblock
\APACrefYearMonthDay{2018}{}{}.
\newblock
{\BBOQ}\APACrefatitle {Alberta wildfire 2016: Apt contribution from anomalous
  planetary wave dynamics} {Alberta wildfire 2016: Apt contribution from
  anomalous planetary wave dynamics}.{\BBCQ}
\newblock
\APACjournalVolNumPages{Scientific Reports}{8}{}{}.
\newblock
\begin{APACrefDOI} \doi{10.1038/s41598-018-30812-z} \end{APACrefDOI}
\PrintBackRefs{\CurrentBib}

\bibitem [\protect \citeauthoryear {%
Petoukhov%
\ \protect \BOthers {.}}{%
Petoukhov%
\ \protect \BOthers {.}}{%
{\protect \APACyear {2016}}%
}]{%
Petoukhov_2016}
\APACinsertmetastar {%
Petoukhov_2016}%
\begin{APACrefauthors}%
Petoukhov, V.%
, Petri, S.%
, Rahmstorf, S.%
, Coumou, D.%
, Kornhuber, K.%
\BCBL {}\ \BBA {} Schellnhuber, H\BPBI J.%
\end{APACrefauthors}%
\unskip\
\newblock
\APACrefYearMonthDay{2016}{}{}.
\newblock
{\BBOQ}\APACrefatitle {Role of quasi-resonant planetary wave dynamics in recent
  boreal spring-to-autumn extreme events} {Role of quasi-resonant planetary
  wave dynamics in recent boreal spring-to-autumn extreme events}.{\BBCQ}
\newblock
\APACjournalVolNumPages{Proc Natl Acad Sci USA}{113}{25}{6862–6867}.
\PrintBackRefs{\CurrentBib}

\bibitem [\protect \citeauthoryear {%
Petoukhov%
, Rahmstorf%
, Petri%
\BCBL {}\ \BBA {} Schellnhuber%
}{%
Petoukhov%
\ \protect \BOthers {.}}{%
{\protect \APACyear {2013}}%
}]{%
Petoukhov_2013}
\APACinsertmetastar {%
Petoukhov_2013}%
\begin{APACrefauthors}%
Petoukhov, V.%
, Rahmstorf, S.%
, Petri, S.%
\BCBL {}\ \BBA {} Schellnhuber, H\BPBI J.%
\end{APACrefauthors}%
\unskip\
\newblock
\APACrefYearMonthDay{2013}{}{}.
\newblock
{\BBOQ}\APACrefatitle {Quasiresonant amplification of planetary waves and
  recent Northern Hemisphere weather extremes} {Quasiresonant amplification of
  planetary waves and recent northern hemisphere weather extremes}.{\BBCQ}
\newblock
\APACjournalVolNumPages{Proc Natl Acad Sci USA}{110}{4}{5336–5341}.
\PrintBackRefs{\CurrentBib}

\bibitem [\protect \citeauthoryear {%
Pfleiderer%
, Schleussner%
\BCBL {}\ \BBA {} Kornhuber%
}{%
Pfleiderer%
\ \protect \BOthers {.}}{%
{\protect \APACyear {2019}}%
}]{%
Pfleiderer&al2019}
\APACinsertmetastar {%
Pfleiderer&al2019}%
\begin{APACrefauthors}%
Pfleiderer, P.%
, Schleussner, C.%
\BCBL {}\ \BBA {} Kornhuber, K\BPBI e\BPBI a.%
\end{APACrefauthors}%
\unskip\
\newblock
\APACrefYearMonthDay{2019}{}{}.
\newblock
{\BBOQ}\APACrefatitle {Summer weather becomes more persistent in a 2 C world.}
  {Summer weather becomes more persistent in a 2 c world.}{\BBCQ}
\newblock
\APACjournalVolNumPages{Nat. Clim. Chang.}{9}{}{666-671}.
\newblock
\begin{APACrefDOI} \doi{10.1038/s41558-019-0555-0} \end{APACrefDOI}
\PrintBackRefs{\CurrentBib}

\bibitem [\protect \citeauthoryear {%
Plotkin%
, Webber%
, O'Neill%
, Weare%
\BCBL {}\ \BBA {} Abbot%
}{%
Plotkin%
\ \protect \BOthers {.}}{%
{\protect \APACyear {2019}}%
}]{%
plotkin2019maximizing}
\APACinsertmetastar {%
plotkin2019maximizing}%
\begin{APACrefauthors}%
Plotkin, D\BPBI A.%
, Webber, R\BPBI J.%
, O'Neill, M\BPBI E.%
, Weare, J.%
\BCBL {}\ \BBA {} Abbot, D\BPBI S.%
\end{APACrefauthors}%
\unskip\
\newblock
\APACrefYearMonthDay{2019}{}{}.
\newblock
{\BBOQ}\APACrefatitle {Maximizing simulated tropical cyclone intensity with
  action minimization} {Maximizing simulated tropical cyclone intensity with
  action minimization}.{\BBCQ}
\newblock
\APACjournalVolNumPages{Journal of Advances in Modeling Earth
  Systems}{11}{4}{863--891}.
\PrintBackRefs{\CurrentBib}

\bibitem [\protect \citeauthoryear {%
Ragone%
\ \BBA {} Bouchet%
}{%
Ragone%
\ \BBA {} Bouchet%
}{%
{\protect \APACyear {2020}}%
}]{%
Ragone_Bouchet2020}
\APACinsertmetastar {%
Ragone_Bouchet2020}%
\begin{APACrefauthors}%
Ragone, F.%
\BCBT {}\ \BBA {} Bouchet, F.%
\end{APACrefauthors}%
\unskip\
\newblock
\APACrefYearMonthDay{2020}{}{}.
\newblock
{\BBOQ}\APACrefatitle {Computation of Extreme Values of Time Averaged
  Observables in Climate Models with Large Deviation Techniques} {Computation
  of extreme values of time averaged observables in climate models with large
  deviation techniques}.{\BBCQ}
\newblock
\APACjournalVolNumPages{J Stat Phys}{179}{}{1637–1665}.
\newblock
\begin{APACrefDOI} \doi{10.1007/s10955-019-02429-7} \end{APACrefDOI}
\PrintBackRefs{\CurrentBib}

\bibitem [\protect \citeauthoryear {%
Ragone%
, Wouters%
\BCBL {}\ \BBA {} Bouchet%
}{%
Ragone%
\ \protect \BOthers {.}}{%
{\protect \APACyear {2018}}%
}]{%
Ragone&al2018}
\APACinsertmetastar {%
Ragone&al2018}%
\begin{APACrefauthors}%
Ragone, F.%
, Wouters, J.%
\BCBL {}\ \BBA {} Bouchet, F.%
\end{APACrefauthors}%
\unskip\
\newblock
\APACrefYearMonthDay{2018}{}{}.
\newblock
{\BBOQ}\APACrefatitle {Computation of extreme heat waves in climate models
  using a large deviation algorithm} {Computation of extreme heat waves in
  climate models using a large deviation algorithm}.{\BBCQ}
\newblock
\APACjournalVolNumPages{Proceedings of the National Academy of
  Sciences}{115}{1}{24-29}.
\newblock
\begin{APACrefDOI} \doi{10.1073/pnas.1712645115} \end{APACrefDOI}
\PrintBackRefs{\CurrentBib}

\bibitem [\protect \citeauthoryear {%
Russo%
, Sillmann%
\BCBL {}\ \BBA {} Fischer%
}{%
Russo%
\ \protect \BOthers {.}}{%
{\protect \APACyear {2015}}%
}]{%
Russo_2015}
\APACinsertmetastar {%
Russo_2015}%
\begin{APACrefauthors}%
Russo, S.%
, Sillmann, J.%
\BCBL {}\ \BBA {} Fischer, E\BPBI M.%
\end{APACrefauthors}%
\unskip\
\newblock
\APACrefYearMonthDay{2015}{}{}.
\newblock
{\BBOQ}\APACrefatitle {Top ten European heatwaves since 1950 and their
  occurrence in the coming decades} {Top ten european heatwaves since 1950 and
  their occurrence in the coming decades}.{\BBCQ}
\newblock
\APACjournalVolNumPages{Environmental Research Letters}{10}{12}{124003}.
\newblock
\begin{APACrefDOI} \doi{10.1088/1748-9326/10/12/124003} \end{APACrefDOI}
\PrintBackRefs{\CurrentBib}

\bibitem [\protect \citeauthoryear {%
Schubert%
, Wang%
, Koster%
, Suarez%
\BCBL {}\ \BBA {} Groisman%
}{%
Schubert%
\ \protect \BOthers {.}}{%
{\protect \APACyear {2014}}%
}]{%
schubert_northern_2014}
\APACinsertmetastar {%
schubert_northern_2014}%
\begin{APACrefauthors}%
Schubert, S.%
, Wang, H.%
, Koster, R.%
, Suarez, M.%
\BCBL {}\ \BBA {} Groisman, P.%
\end{APACrefauthors}%
\unskip\
\newblock
\APACrefYearMonthDay{2014}{}{}.
\newblock
{\BBOQ}\APACrefatitle {Northern {{Eurasian Heat Waves}} and {{Droughts}}}
  {Northern {{Eurasian Heat Waves}} and {{Droughts}}}.{\BBCQ}
\newblock
\APACjournalVolNumPages{Journal of Climate}{27}{9}{3169-3207}.
\newblock
\begin{APACrefDOI} \doi{10.1175/JCLI-D-13-00360.1} \end{APACrefDOI}
\PrintBackRefs{\CurrentBib}

\bibitem [\protect \citeauthoryear {%
Schubert%
, Wang%
\BCBL {}\ \BBA {} Suarez%
}{%
Schubert%
\ \protect \BOthers {.}}{%
{\protect \APACyear {2011}}%
}]{%
Schubert_2011}
\APACinsertmetastar {%
Schubert_2011}%
\begin{APACrefauthors}%
Schubert, S.%
, Wang, H.%
\BCBL {}\ \BBA {} Suarez, M.%
\end{APACrefauthors}%
\unskip\
\newblock
\APACrefYearMonthDay{2011}{}{}.
\newblock
{\BBOQ}\APACrefatitle {Warm season subseasonal variability and climate extremes
  in the northern hemisphere: the role of stationary Rossby waves} {Warm season
  subseasonal variability and climate extremes in the northern hemisphere: the
  role of stationary rossby waves}.{\BBCQ}
\newblock
\APACjournalVolNumPages{J Clim}{24}{8}{4773-4792}.
\PrintBackRefs{\CurrentBib}

\bibitem [\protect \citeauthoryear {%
St\'efanon%
, D'Andrea%
\BCBL {}\ \BBA {} Drobinski%
}{%
St\'efanon%
\ \protect \BOthers {.}}{%
{\protect \APACyear {2012}}%
}]{%
Stefanon_2012}
\APACinsertmetastar {%
Stefanon_2012}%
\begin{APACrefauthors}%
St\'efanon, M.%
, D'Andrea, F.%
\BCBL {}\ \BBA {} Drobinski, P.%
\end{APACrefauthors}%
\unskip\
\newblock
\APACrefYearMonthDay{2012}{}{}.
\newblock
{\BBOQ}\APACrefatitle {Heatwave classification over Europe and the
  Mediterranean region} {Heatwave classification over europe and the
  mediterranean region}.{\BBCQ}
\newblock
\APACjournalVolNumPages{Environmental Research Letters}{7}{}{014023}.
\newblock
\begin{APACrefDOI} \doi{10.1088/1748-9326/7/1/014023} \end{APACrefDOI}
\PrintBackRefs{\CurrentBib}

\bibitem [\protect \citeauthoryear {%
Suarez-Gutierrez%
, Li%
, Müller%
\BCBL {}\ \BBA {} Marotzke%
}{%
Suarez-Gutierrez%
\ \protect \BOthers {.}}{%
{\protect \APACyear {2018}}%
}]{%
Suarez_Gutierrez_2018}
\APACinsertmetastar {%
Suarez_Gutierrez_2018}%
\begin{APACrefauthors}%
Suarez-Gutierrez, L.%
, Li, C.%
, Müller, W\BPBI A.%
\BCBL {}\ \BBA {} Marotzke, J.%
\end{APACrefauthors}%
\unskip\
\newblock
\APACrefYearMonthDay{2018}{}{}.
\newblock
{\BBOQ}\APACrefatitle {Internal variability in European summer temperatures at
  1.5 C and 2 C of global warming} {Internal variability in european summer
  temperatures at 1.5 c and 2 c of global warming}.{\BBCQ}
\newblock
\APACjournalVolNumPages{Environmental Research Letters}{13}{6}{064026}.
\newblock
\begin{APACrefDOI} \doi{10.1088/1748-9326/aaba58} \end{APACrefDOI}
\PrintBackRefs{\CurrentBib}

\bibitem [\protect \citeauthoryear {%
Teng%
, Branstator%
, Wang%
, Meehl%
\BCBL {}\ \BBA {} Washington%
}{%
Teng%
\ \protect \BOthers {.}}{%
{\protect \APACyear {2013}}%
}]{%
Branstator_2013}
\APACinsertmetastar {%
Branstator_2013}%
\begin{APACrefauthors}%
Teng, H.%
, Branstator, G.%
, Wang, H.%
, Meehl, G.%
\BCBL {}\ \BBA {} Washington, W.%
\end{APACrefauthors}%
\unskip\
\newblock
\APACrefYearMonthDay{2013}{10}{}.
\newblock
{\BBOQ}\APACrefatitle {Probability of US heat waves affected by a subseasonal
  planetary wave pattern} {Probability of us heat waves affected by a
  subseasonal planetary wave pattern}.{\BBCQ}
\newblock
\APACjournalVolNumPages{Nature Geoscience}{6}{}{1056-1061}.
\newblock
\begin{APACrefDOI} \doi{10.1038/ngeo1988} \end{APACrefDOI}
\PrintBackRefs{\CurrentBib}

\bibitem [\protect \citeauthoryear {%
Trenberth%
}{%
Trenberth%
}{%
{\protect \APACyear {2012}}%
}]{%
Trenberth2012}
\APACinsertmetastar {%
Trenberth2012}%
\begin{APACrefauthors}%
Trenberth, K.%
\end{APACrefauthors}%
\unskip\
\newblock
\APACrefYearMonthDay{2012}{}{}.
\newblock
{\BBOQ}\APACrefatitle {Framing the way to relate climate extremes to climate
  change} {Framing the way to relate climate extremes to climate
  change}.{\BBCQ}
\newblock
\APACjournalVolNumPages{Clim. Change}{115}{}{283-290}.
\PrintBackRefs{\CurrentBib}

\bibitem [\protect \citeauthoryear {%
Vautard%
\ \protect \BOthers {.}}{%
Vautard%
\ \protect \BOthers {.}}{%
{\protect \APACyear {2007}}%
}]{%
Vautard&al2007}
\APACinsertmetastar {%
Vautard&al2007}%
\begin{APACrefauthors}%
Vautard, R.%
, Yiou, P.%
, D'Andrea, F.%
, de Noblet, N.%
, Viovy, N.%
, Cassou, C.%
\BDBL {}Fan, Y.%
\end{APACrefauthors}%
\unskip\
\newblock
\APACrefYearMonthDay{2007}{}{}.
\newblock
{\BBOQ}\APACrefatitle {Summertime European heat and drought waves induced by
  wintertime Mediterranean rainfall deficit} {Summertime european heat and
  drought waves induced by wintertime mediterranean rainfall deficit}.{\BBCQ}
\newblock
\APACjournalVolNumPages{Geophysical Research Letters}{34}{7}{}.
\newblock
\begin{APACrefDOI} \doi{10.1029/2006GL028001} \end{APACrefDOI}
\PrintBackRefs{\CurrentBib}

\bibitem [\protect \citeauthoryear {%
Webber%
, Plotkin%
, O’Neill%
, Abbot%
\BCBL {}\ \BBA {} Weare%
}{%
Webber%
\ \protect \BOthers {.}}{%
{\protect \APACyear {2019}}%
}]{%
webber_practical_2019}
\APACinsertmetastar {%
webber_practical_2019}%
\begin{APACrefauthors}%
Webber, R\BPBI J.%
, Plotkin, D\BPBI A.%
, O’Neill, M\BPBI E.%
, Abbot, D\BPBI S.%
\BCBL {}\ \BBA {} Weare, J.%
\end{APACrefauthors}%
\unskip\
\newblock
\APACrefYearMonthDay{2019}{}{}.
\newblock
{\BBOQ}\APACrefatitle {Practical rare event sampling for extreme mesoscale
  weather} {Practical rare event sampling for extreme mesoscale
  weather}.{\BBCQ}
\newblock
\APACjournalVolNumPages{Chaos: An Interdisciplinary Journal of Nonlinear
  Science}{29}{5}{053109}.
\newblock
\begin{APACrefDOI} \doi{10.1063/1.5081461} \end{APACrefDOI}
\PrintBackRefs{\CurrentBib}

\bibitem [\protect \citeauthoryear {%
Zampieri%
\ \protect \BOthers {.}}{%
Zampieri%
\ \protect \BOthers {.}}{%
{\protect \APACyear {2014}}%
}]{%
Zampieri&al2014}
\APACinsertmetastar {%
Zampieri&al2014}%
\begin{APACrefauthors}%
Zampieri, M.%
, D'Andrea, F.%
, Vautard, R.%
, Ciais, P.%
, de Noblet-Ducoudr{é}, N.%
\BCBL {}\ \BBA {} Yiou, P.%
\end{APACrefauthors}%
\unskip\
\newblock
\APACrefYearMonthDay{2014}{}{}.
\newblock
{\BBOQ}\APACrefatitle {Hot {E}uropean Summers and the Role of Soil Moisture in
  the Propagation of Mediterranean Drought} {Hot {E}uropean summers and the
  role of soil moisture in the propagation of mediterranean drought}.{\BBCQ}
\newblock
\APACjournalVolNumPages{J. Climate}{22}{}{4747-4758}.
\newblock
\begin{APACrefDOI} \doi{10.1175/2009JCLI2568.1} \end{APACrefDOI}
\PrintBackRefs{\CurrentBib}

\end{thebibliography}


%
%

%
%
%
%
%

\end{document}